\documentstyle[epsf,psfig,amsmath,amssymb]{article}
\begin{document}

\begin{center}
{\large\bf Preparation of W, GHZ, and two-qutrit states using bimodal cavities}

ASOKA BISWAS and G. S. AGARWAL

Physical Research Laboratory, Navrangpura, Ahmedabad - 380 009, India;
e-mail: asoka@prl.ernet.in

\date{\today}
\end{center}

{\bf Abstract :} {\small We show how one can prepare three-qubit entangled states like W-states, 
Greenberger-Horne-Zeilinger states as well as two-qutrit entangled states using the multiatom two-mode entanglement.
We propose a technique of preparing such a multi-particle entanglement using stimulated Raman
adiabatic passage. We consider a collection of three-level atoms in $\Lambda$ configuration simultaneously 
interacting with a resonant two-mode cavity for this purpose. Our approach permits a variety of multiparticle extensions.}

\section{\label{sec:intro}Introduction}

\indent

Entanglement between two qubits can be produced only if there is some kind of 
interaction between the qubits \cite{chuang,divincenzo,unanyan,bose}. 
This could be the exchange interaction 
between the two spins or the dipole-dipole interaction between, say, two 
atoms. The entanglement produced by direct interaction has been extensively 
studied \cite{dd}. However the direct interaction falls very rapidly as the distance 
between qubits increases. It turns our that in such situations one could 
produce entanglement using cavity fields \cite{haroche:rmp}. The non-interacting atoms get coupled 
because of the interaction with a common quantized field. Agarwal 
{\it et al.\/} had shown earlier \cite{agarwal:97}, how coherently prepared atoms, on 
interaction with the vacuum of the cavity field produce mesoscopic 
superpositions which exhibit the same type of entanglement character as the 
GHZ states. There have been many proposals for producing entanglement using 
cavities \cite{gsa:solano,kim:knight}. Haroche and his coworkers have produced variety of entanglement using cavity QED based 
schemes. In this paper we show how bimodal cavities can be used for producing 
{\it multiatom and bimodal field entanglement} and how this entanglement can 
be used to produce a variety of entangled states such as W-states \cite{wstate}, GHZ states \cite{ghz:state} 
and two-qutrit states \cite{twoqutrit}. We consider the interaction of $N$ three-level atoms 
in $\Lambda$ configuration interacting with a bimodal cavity for this purpose.
The entangled states thus prepared are comprised of ground states of the 
three-level atoms.

We note that W-states, GHZ states and  a host of other types of 
multimode states of the radiation field have been realized using the fields
produced in the process of downconversion by using appropriate detection 
schemes \cite{weinfurter,zeilinger}. Further entanglement between atoms has also been realized by 
suitable projection schemes \cite{polzik:nature,zoller:cabrillo,gsa:zanthier}. 
Detection procedures have been found to be useful in implementing nondeterministic logic gates also
\cite{franson}. 

The structure of this 
paper is as follows. In Sec.~\ref{sec:basic}, we describe the generalized 
model, in which two atoms interact with two-mode quantized field.  
We show how a two-atom two-mode entangled state is produced 
during this process. In Sec.~\ref{sec:epr}, we explore the
possibility of preparation of different kinds of entangled states either by adiabatic technique or by
the method of projection from the prepared state.  
In Sec.~IV, we propose the extension of 
the three-particle entangled states to those prepared in the process involving 
 $N$ atoms. 

\begin{figure}[h]
\epsfxsize 12cm
\centerline{
\epsfbox{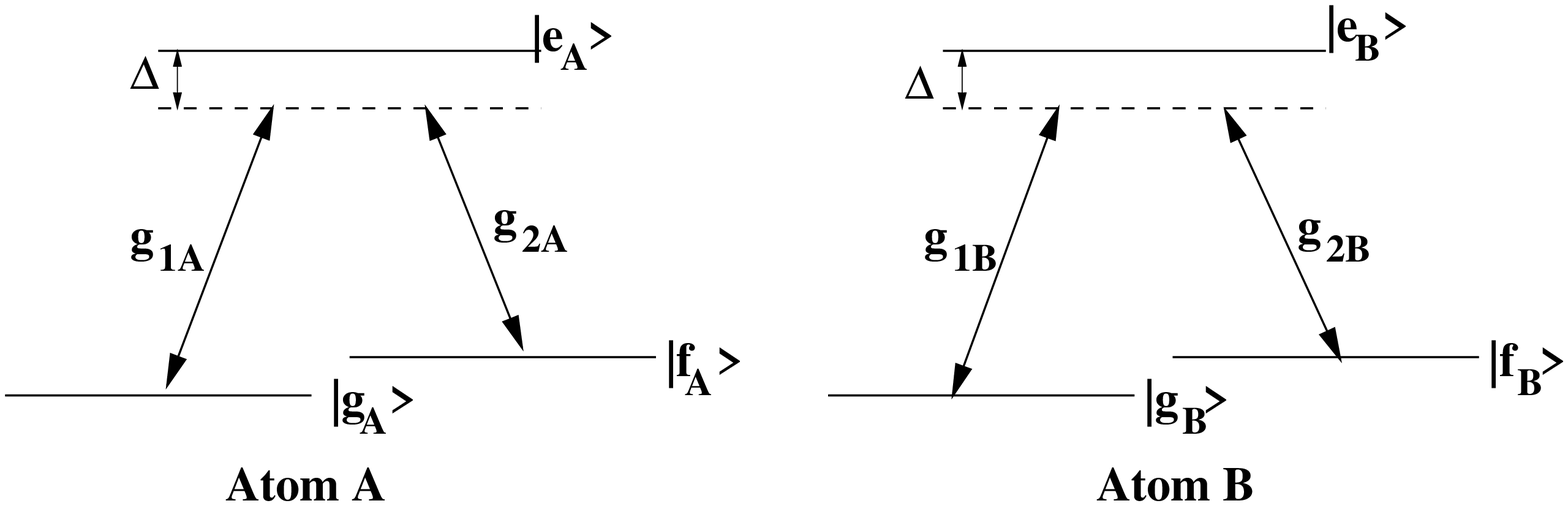}}
\caption{\label{fig1}{\small Level diagram  of two three-level atoms in
$\Lambda$-configuration, interacting with two cavity modes defined by annihilation operators
$a$ and $b$. $g_{1k}$ and $g_{2k}$ $(k=$ A,B)
are the atom-cavity coupling terms for the $k$-th atom. $\Delta$ is the common one-photon
detuning of the fields.}}
\end{figure}

\section{\label{sec:basic}Preparation of a two-atom two-mode entangled state} 

\indent

We consider two three-level atoms (A and B) with relevant energy levels in
$\Lambda$-configuration (see Fig.~\ref{fig1}) interacting with a two-mode high quality 
optical cavity. The specified annihilation operators for the 
cavity modes are $a$ and $b$. The atoms are interacting with the cavity mode $a$ in 
$|e\rangle \leftrightarrow |g\rangle$ transition and with the mode $b$ in 
$|e\rangle \leftrightarrow |f\rangle$ transition. 

The Hamiltonian for the system under rotating wave approximation can be written as 
\begin{equation}
\label{hamil1}H=\hbar(\omega_1 a^\dag a+\omega_2 b^\dag b)+\hbar\sum_{k=A,B}\left[\omega_{e_kg_k}|e_k\rangle\langle e_k|+\omega_{f_kg_k}|f_k\rangle\langle f_k|+\left\{g_{1k}|e_k\rangle\langle g_k|a+g_{2k}|e_k\rangle\langle f_k|b+\textrm{h.c.}\right\}\right]\;,
\end{equation}
where, $\omega_{l_km_k}$ is the atomic transition frequency 
between the levels $|l_k\rangle$ and $|m_k\rangle$, 
$\omega_j$ $(j=1,2)$ is the respective frequency of the cavity modes $a$ and 
$b$, $g_{jk}$ $(j=1,2)$ provides the atom-cavity coupling. We assume $g_{jk}$'s to be real and function of time.

We start with the initial state $|\psi_i\rangle =|g_A,g_B,n,\mu\rangle$, where
$n$ and $\mu$ are the initial numbers of photons in the
cavity modes $a$ and $b$, respectively and the two atoms are in $|g\rangle$ 
state.
The state of the system can be expanded in terms of the relevant basis 
states in the following way: 
\begin{eqnarray}
|\psi(t)\rangle &=& c_1|g_A,g_B,n,\mu\rangle +c_2|g_A,e_B,n-1,\mu\rangle +c_3|g_A,f_B,n-1,\mu +1\rangle\nonumber\\
&& +c_4|e_A,f_B,n-2,\mu +1\rangle +c_5|e_A,e_B,n-2,\mu\rangle +c_6|f_A,f_B,n-2,\mu +2\rangle \nonumber \\
&& +c_7|f_A,e_B,n-2,\mu +1\rangle +c_8|f_A,g_B,n-1,\mu +1\rangle +c_9|e_A,g_B,n-1,\mu\rangle \;.\label{wavefunc}
\end{eqnarray}
From the Schr{\"o}dinger equation we find the  equations for the 
probability amplitudes which can be solved numerically.
 
 Under the two-photon resonance in each atom as shown in the Fig.~\ref{fig1}, 
we find that one of the eigenvalues of the Hamiltonian in the interaction picture is zero. The corresponding eigenstate is found to be
\begin{eqnarray}
|\psi_0\rangle &=& \frac{1}{P}\left[g_{2A}g_{2B}\sqrt{(\mu +1)(\mu +2)}|g_A,g_B,n,\mu\rangle +g_{1A}g_{1B}\sqrt{n(n-1)}|f_A,f_B,n-2,\mu +2\rangle\right.\nonumber\\
&& \left.-g_{1B}g_{2A}\sqrt{n(\mu +2)}|g_A,f_B,n-1,\mu +1\rangle-g_{1A}g_{2B}\sqrt{n(\mu +2)}|f_A,g_B,n-1,\mu +1\rangle)\right]\;,\nonumber\\
\label{dark}
\end{eqnarray}
where, 
\begin{equation}
P=[g_{2A}^2g_{2B}^2(\mu +1)(\mu +2)+g_{1A}^2g_{1B}^2 n(n-1)+(g_{1B}^2g_{2A}^2+g_{1A}^2g_{2B}^2) n(\mu +2)]^{1/2}\;.
\label{norm}
\end{equation}
Clearly, this state is a two-atom two-mode (four-particle) entangled state. 
We note that recently there have been a few demonstrations of the preparation of four-particle entangled state.
For example, entangled states of four photons have been prepared using a single \cite{weinfurter} or two \cite{zeilinger} parametric down-converter sources, 
whereas, entangled states of four ions have been prepared using a single laser pulse by Sackett {\it et al.\/} \cite{sackett}. 
We next show how different classes of entangled states can be derived from the 
 state (\ref{dark}).

\begin{figure}
\centerline{
\begin{tabular}{cc}
\psfig{figure=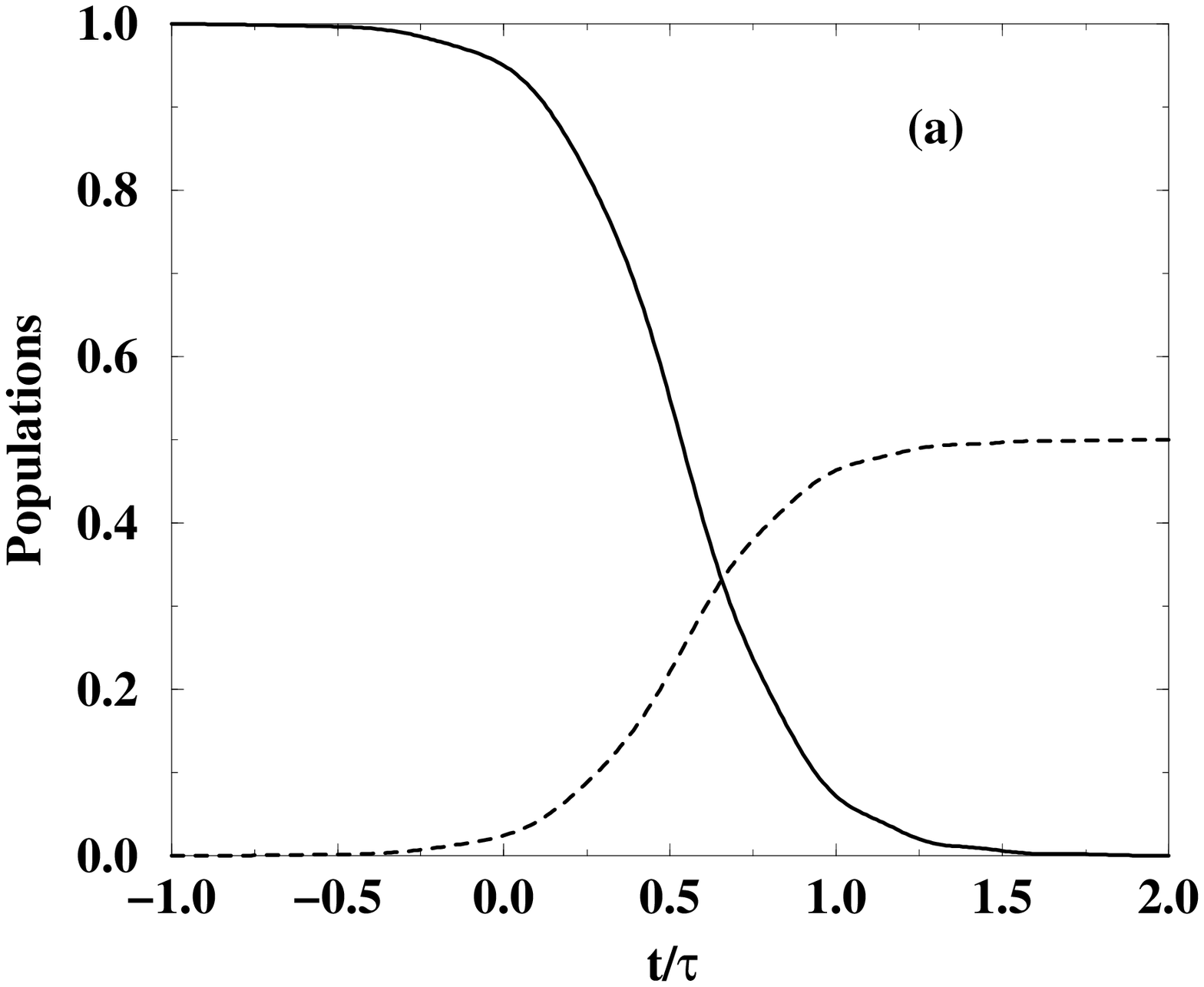,height=6.0cm}&
\psfig{figure=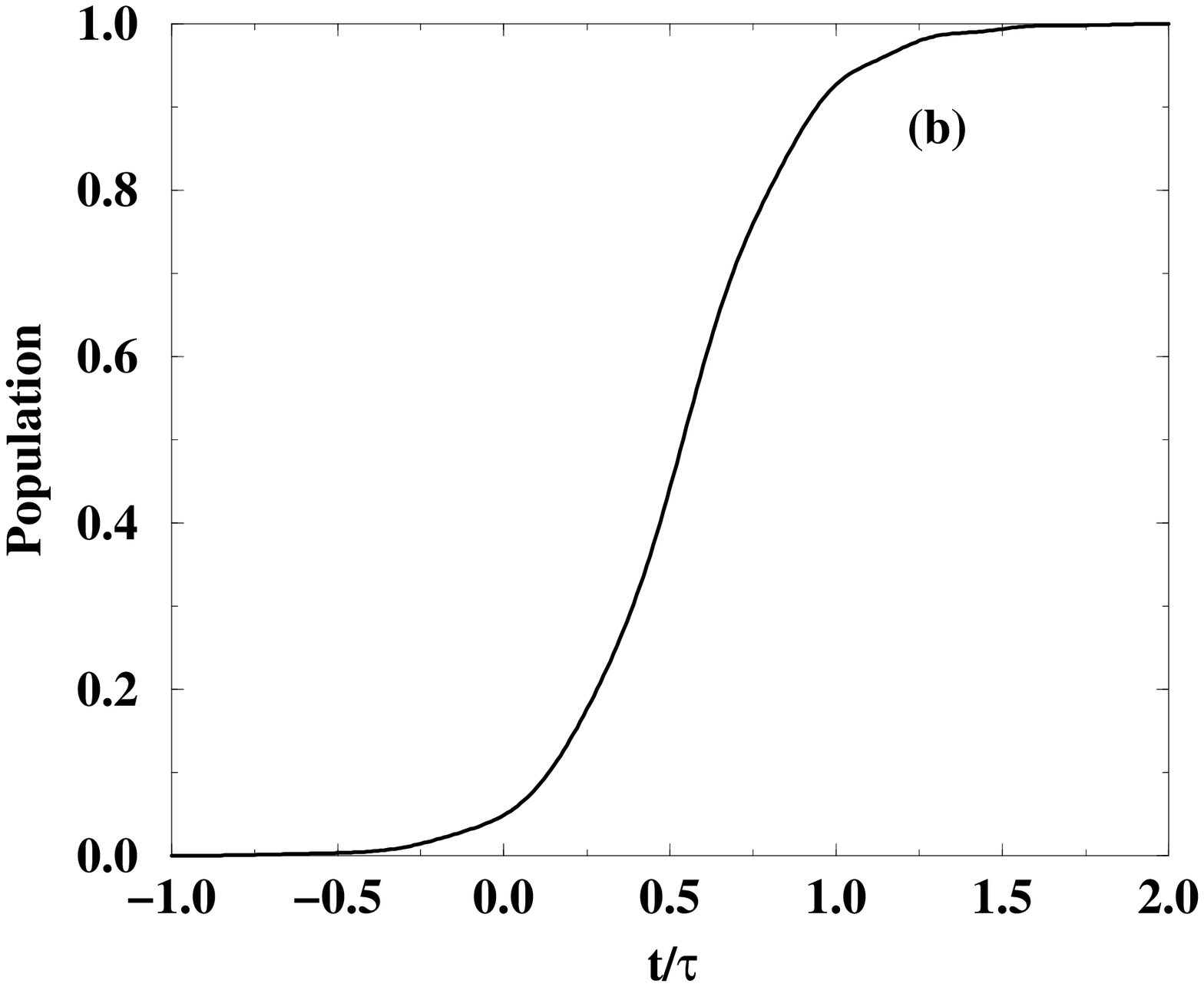,height=6.0cm}
\end{tabular}}
\caption{\label{fig2}{\small (a) Time evolution of populations in the relevant energy levels for $n=1$, $\mu=0$, $\Delta\tau=0$, $g_{j0}\tau=15$ ($j=1,2$), and $T=4\tau/3$.
(b) Time evolution of the entangled state
(\ref{epreq}) for $n=1$. The other parameters are same 
as in (a).}}
\end{figure}

\section{\label{sec:epr}Preparation of different classes of entangled states}

\indent

We will now discuss how different classes of entangled states can be prepared either in a 
deterministic way using delayed pulses or in probabilistic way by detecting the atoms
or cavity in different basis.

\subsection{\label{sec:delay}Deterministic Entanglement}
\subsubsection{Preparation of two-particle maximally entangled states}

\indent

We assume that both the atoms are in $|g\rangle$ state and there is initially one photon present in $a$ mode. Thus the system
initially is prepared in the state $|g_A,g_B,1,\mu\rangle$. Note that a Fock state
in a particular mode can be prepared by sending suitably chosen excited atoms through the cavity \cite{walther}. We now choose the
time-dependence of the cavity couplings as 
\begin{equation}
g_{2k}=g_{20} e^{-t^2/\tau^2};~~~g_{1k}=g_{10} e^{-(t-T)^2/\tau^2}\;,~k=A, B\;,
\label{pulseseq}
\end{equation}
where, $g_{j0}$ $(j=1,2)$ is the respective amplitude of the pulses, $\tau$ and $T$ are the width and the time-separation
respectively of the pulses. Note that the pulses are applied in counterintuitive
sequence in the sense that the pulse on the Stokes transition $|e\rangle\leftrightarrow |f\rangle$ acts before the pulse on the transition $|e\rangle\leftrightarrow |g\rangle$.
Here we should mention that the idea of counterintuitive sequence of pulses 
have been first incorporated in the process of population transfer between 
two dipole-forbidden ground levels of a $\Lambda$ system (Stimulated Raman adiabatic passage : STIRAP) \cite{stirap}. In that process two classical pulses have
been employed in two-photon resonance condition. Later, this idea had been
extended by Parkins {\it et al.\/} \cite{parkins} to single-mode cavity to 
prepare superposition of Fock states. Rempe and his coworkers have verified 
experimentally the stimulated Raman scattering using adibatic passage technique
in a high-finesse optical cavity \cite{rempe}. In the context of cavity QED 
and two-level atoms, Stenholm and
coworkers have studied adiabatic state preparation in the single-mode
case \cite{sten1} and the photon statistics anomalies
in adiabatic transfer between two cavity modes \cite{sten2}.

In the present model, due to counterintuitive sequence of the pulses (\ref{pulseseq}), the atom-cavity system now follows the
evolution of the null eigenstate (\ref{dark}) which is an adiabatic state \cite{stirap}.
From Eq.~(\ref{dark}), we see that, in the presence of the above pulses the population will be adiabatically transferred from the
state $|g_A,g_B,1,\mu\rangle$ to the states  $|g_A,f_B,0,\mu+1\rangle$ and $|f_A,g_B,0,\mu+1\rangle$ at long time. 
Thus, an equal superposition of $|g_A,f_B,0,\mu+1\rangle$ and
$|f_A,g_B,0,\mu+1\rangle$ is prepared at the end of the evolution [see Fig.~\ref{fig2}(a)]. The final state of the system can be written as
\begin{equation}
\frac{1}{\sqrt{2}}(|g_A,f_B,0,\mu+1\rangle+|f_A,g_B,0,\mu+1\rangle)\;.
\end{equation}
Clearly this state is disentangled in the Hilbert space of atoms and cavity modes. The
atoms are prepared in the following two-particle entangled state
\begin{equation}
\label{epreq}\frac{1}{\sqrt{2}}(|g_A,f_B\rangle+|f_A,g_B\rangle)\;,
\end{equation}
which is the symmetric combination of atomic ground  states and hence is radiatively stable.
We show the temporal build up of this state in Fig.~\ref{fig2}(b). 
We should mention here that the both symmetric and anti-symmetric entangled 
states comprised of the ground states of two three-level atoms can also be 
prepared by the aid of resonant dipole-dipole interaction between them 
\cite{dd}. However in that case entanglement decreases very fast as the 
interatomic distance increases beyond dimensions of the order of a wavelength. 
Further,
Unanyan {\it et al.\/} have  proposed the use of adiabatic passage technique to prepare 
states of the form (\ref{epreq}) \cite{unanyan}. An important ingredient in the 
work of Ref.~\cite{unanyan} is the {\it exchange interaction}. 

\begin{figure}
\centerline{
\begin{tabular}{cc}
\psfig{figure=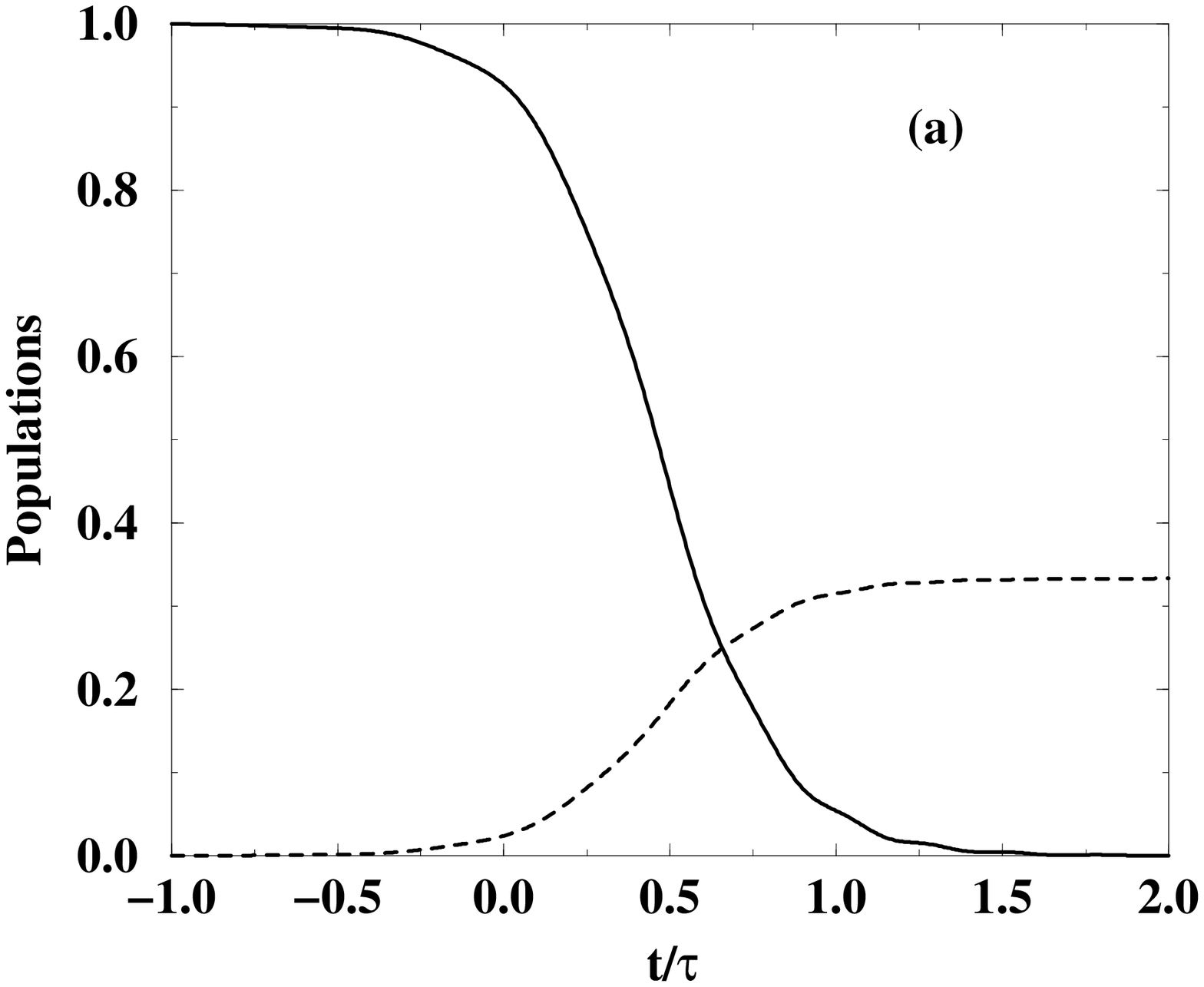,height=6.0cm}&
\psfig{figure=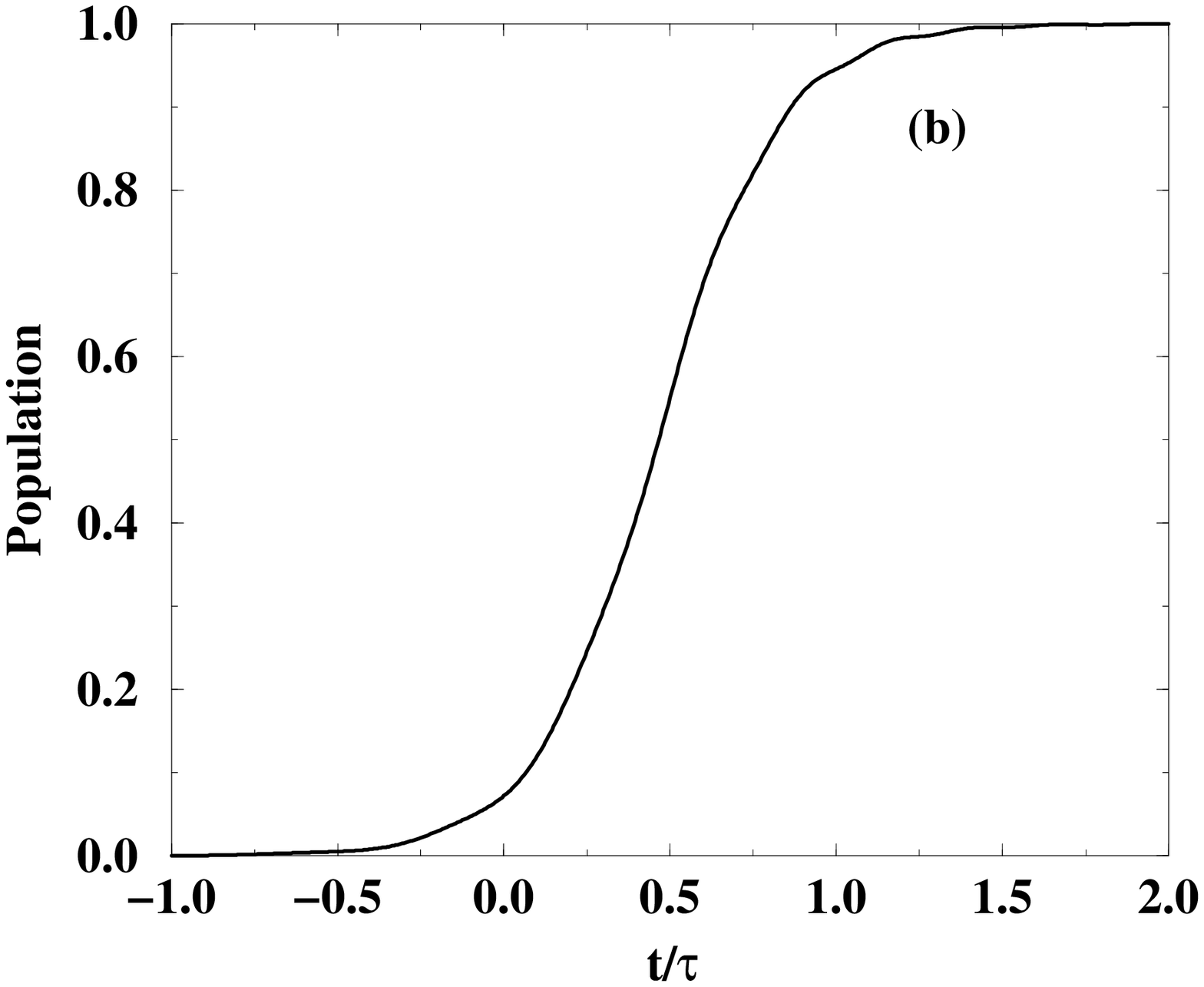,height=6.0cm}
\end{tabular}}
\caption{\label{fig4}{\small (a)Time evolution of the populations in the relevant energy levels for $n=1$, $\mu=0$, $\Delta\tau=0$,
 $g_{j0}\tau=15$ ($j=1,2$), and $T=4\tau/3$. Here the solid line represents the population of the level $|g_A,g_B,g_C,1,\mu\rangle$ 
 and the dashed line represents the populations of the levels $|g_A,g_B,f_C,0,1\rangle$, $|g_A,f_B,g_C,0,1\rangle$, and 
 $|f_A,g_B,g_C,0,1\rangle$. (b) Time evolution of the three-atom W state (\ref{stateW})
 for the parameters same as in (a).}}
\end{figure}
 
\subsubsection{\label{sec:epr1}Preparation of a system in W-state}

\indent

The W-state is a tripartite entangled state in which either of the qubit remains in the
excited state while others remain in the ground state. The general form of
this state is
\begin{equation}
\frac{1}{\sqrt{3}}(|0,0,1\rangle+|0,1,0\rangle+|1,0,0\rangle)\;.
\end{equation}
The interesting property of this state is that it retains bipartite entanglement
when one of
the qubits is traced out. In fact, the entanglement in W-state has the highest
degree of endurance against qubit loss \cite{wstate}.

This kind of state can also be prepared by
considering a system of three atoms in state $|g\rangle$ 
passing through  the two-mode cavity with a
single photon initially in the $a$-mode. Thus the system is initially 
prepared in
a state $|g_A,g_B,g_C,1,\mu\rangle$. 
If one uses the pulse sequence (\ref{pulseseq}), where $k=$A, B, C, then
the population will be adiabatically transferred to the state $|g_A,g_B,f_C,0,\mu +1\rangle$,
$|g_A,f_B,g_C,0,\mu +1\rangle$, and $|f_A,g_B,g_C,0,\mu +1\rangle$ at long time through 
a corresponding three-atom two-mode dark state. The temporal evolution of the populations
in these states has been shown in Fig.~\ref{fig4}(a). Clearly, at the end of the evolution,
the atoms are prepared in the following three-particle entangled state:
\begin{equation}
\label{stateW}\frac{1}{\sqrt{3}}(|g_A,g_B,f_C\rangle+|g_A,f_B,g_C\rangle+|f_A,g_B,g_C\rangle)\;,
\end{equation}
The temporal build up of this state is shown in Fig.~\ref{fig4}(b).
Note that the state (\ref{stateW}) is a W-state in the basis of
three atoms. This state is immune to spontaneous emission as is 
prepared of ground states of the atoms. We note that, recently there is an 
experimental demonstration of three-photon W-state in photon polarization basis 
using the method of spontaneous parametric down conversion \cite{jmo}.

\subsection{Entanglement by Detection} 
\subsubsection{\label{sec:entangle}Preparation of GHZ states}

\indent

We will now discuss how another  kind of tripartite entangled states, namely
GHZ states \cite{ghz:state} can be prepared using the present scheme.
Preparation of three-photon GHZ states has already been experimentally demonstrated \cite{bou}.

We continue to use two atoms interacting
 with a two-mode cavity. Here, we consider the cavity coupling for any mode is same for
both the atoms, i.e., $g_{1A}=g_{1B}=g_1$ and $g_{2A}=g_{2B}=g_2$. We also
assume that the cavity fields are adiabatically turned on so that the system 
initially in the state $|g_A,g_B,n,\mu\rangle$ is prepared in the null eigenstate 
$|\psi_0\rangle$  given by
\begin{equation}
|\psi_0\rangle =\frac{1}{P}\left[\alpha|g_A,g_B,n,\mu\rangle + \beta|f_A,f_B,n-2,\mu +2\rangle +\gamma(|g_A,f_B,n-1,\mu +1\rangle+|f_A,g_B,n-1,\mu +1\rangle\right]\;,
\label{dark1}
\end{equation}
where,
\begin{equation}
\alpha=g_2^2\sqrt{(\mu +1)(\mu +2)}\;,~ \beta=g_1^2\sqrt{n(n-1)}\;,~\gamma=-g_1g_2\sqrt{n(\mu +2)}\;,~\textrm{and}~~P=\sqrt{\alpha^2+\beta^2+2\gamma^2}\;.
\label{norm1}
\end{equation}
Now the cavity pulses are turned off suddenly so  that the system remains in 
the state $|\psi_0\rangle$ without further evolution. The values $g_1$ and 
$g_2$ in (\ref{dark1}) are the same as at the time when the pulses were turned 
off.
We show how the detection of the state 
$|\psi_0\rangle$ in some pre-determined basis helps us to prepare different entangled states.  
  
Firstly if one projects $|\psi_0\rangle$ into the field basis $|n-1,\mu+1\rangle$, 
the system collapses to an EPR state $(|g_A\rangle|f_B\rangle+|f_A\rangle|g_B\rangle)/\sqrt{2}$ 
in two-atom basis, as clear from (\ref{dark1}).
We note that Can {\it et al.\/} \cite{sakir} have prepared similar kind of two-atom maximally
entangled state by detecting the photon in $b$-mode leaking out of the cavity,
as in our case with $n=1$ and $\mu=0$.

Next, a measurement in the basis $|g_B\rangle$ or $|f_B\rangle$ prepares the
system in the following GHZ equivalent states with atom $A$ and the cavity modes:
\begin{eqnarray}
&&\frac{1}{C'}[\alpha|g_A,n,\mu\rangle+\gamma |f_A,n-1,\mu+1\rangle]\nonumber\\
&\textrm{or,}&\frac{1}{C''}[\beta|f_A,n-2,\mu+2\rangle+\gamma|g_A,n-1,\mu+1\rangle]\;,
\end{eqnarray}
where, $C'=(\alpha^2+\gamma^2)^{1/2}$ and $C''=(\beta^2+\gamma^2)^{1/2}$.
Similar kinds of GHZ equivalent states with atom B and cavity modes can be prepared
if we project the state $|\psi_0\rangle$ (\ref{dark1}) either in $|g_A\rangle$ or
in $|f_A\rangle$.
 
We should mention here that the present procedure is probabilistic unlike the case in 
Sec.~\ref{sec:delay}, which was based on delayed pulse
technique and thus was deterministic. We note that the generation of different kinds of entangled states by
detection is a very popular scheme \cite{zeilinger,sakir,tri,lee,milburn,zou,dowling}.

\subsubsection{\label{sec:epr2}Preparation of two-qutrit entangled states}

\indent

We will now discuss how one can prepare the entangled states 
of two qutrits (i.e., quantum particles defined in three-dimensional Hilbert 
space). Recently, two-qutrit entangled states have been found to be much useful
in quantum communication due to 
their high noise resistance \cite{twoqutrit} and in quantum cryptography \cite{peres,vaziri}. 
Zou {\it et al.\/} have proposed a scheme to generate two-qutrit states \cite{zouqutrit}.
We note that Vaziri {\it et al.\/} have demonstrated preparation of two-qutrit 
states for the radiation field using down-conversion photon source \cite{vaziri}.

To prepare such kind of states using our model, we prepare the system in the state (\ref{dark1}) in the way
as discussed in the previous subsection. We can rewrite this state as
\begin{eqnarray}
|\psi_0\rangle &=&\frac{1}{2P}\left[|\psi^+\rangle_A|\psi^+\rangle_B (\alpha|n,\mu\rangle+\beta|n-2,\mu+2\rangle+2\gamma|n-1,\mu+1\rangle)\right.\nonumber\\
&+&|\psi^+\rangle_A|\psi^-\rangle_B(\alpha|n,\mu\rangle-\beta|n-2,\mu+2\rangle)\nonumber\\
&+&|\psi^-\rangle_A|\psi^-\rangle_B (\alpha|n,\mu\rangle+\beta|n-2,\mu+2\rangle-2\gamma|n-1,\mu+1\rangle)\nonumber\\
&+&\left.|\psi^-\rangle_A|\psi^+\rangle_B(\alpha|n,\mu\rangle-\beta|n-2,\mu+2\rangle)\right]\;, 
\end{eqnarray}
where, $|\psi^\pm\rangle_k = (|g\rangle\pm|f\rangle)_k/\sqrt{2}$ $(k=$ A, B).
Clearly, by measuring both the atoms A and B in $|\psi^\pm\rangle$, one can prepare the entangled two-qutrit states 
\begin{equation}
|\Psi_\pm\rangle=\frac{1}{P'}(\alpha|n,\mu\rangle+\beta|n-2,\mu+2\rangle\pm 2\gamma|n-1,\mu+1\rangle)\;,
\end{equation}
where, $P'=[\alpha^2+\beta^2+4\gamma^2]^{1/2}$.
One can easily identify the two cavity modes $a$ and $b$ with the respective basis states $(|n\rangle,|n-1\rangle,|n-2\rangle)$ and $(|\mu\rangle,|\mu+1\rangle,|\mu+2\rangle)$ as two qutrits.
This kind of projection measurement can be done using Ramsey interferometry.

\section{Extension to $N$-particle entanglement}

\indent

In this section, we propose the possible extension of our method to generate 
the multipartite entangled states involving $N$ atoms.
For example, if one prepares $N$ atoms in $|g\rangle$ state and start with a single photon in $a$ mode, then
using the counter-intuitive pulse sequence as discussed in Sec.~\ref{sec:epr1}, the
following $N$-atom W-state can be prepared:
\begin{equation}
\frac{1}{\sqrt{N}}\left[\sum_{k=1}^N\left(\prod_{k'\neq k}|g_{k'}\rangle|f_k\rangle\right)\right]\;.
\end{equation}
We note that $N$-particle W-state can also be prepared using $N$ two-level atoms interacting with 
resonant \cite{statew1} or detuned \cite{lili} single-mode cavities.

Further, there is a possibility of extension of tripartite GHZ entangled states to $N$ particles. One can
prepare a $N$-atom two-mode  dark state equivalent to the two-atom two-mode dark state (\ref{dark1}) in
a similar way as discussed in Sec.~\ref{sec:entangle}. In that case,
one has to prepare all the atoms in $|g\rangle$ state and the cavity with $n$ ($\ge N$) photons in $a$
mode. Thus, the GHZ-equivalent $(N+1)$-partite entangled states
\begin{eqnarray}
&&\frac{1}{C'}\left[\alpha\prod_{k=1}^{N-1}|g_k\rangle |n,\mu\rangle+ \gamma \prod_{k=1}^{N-1}|f_k\rangle|n-N+1,\mu+N-1\rangle\right]\nonumber\\
\textrm{or,}&&\frac{1}{C''}\left[\beta\prod_{k=1}^{N-1}|f_k\rangle|n-N,\mu+N\rangle+\gamma\prod_{k=1}^{N-1}|g_k\rangle|n-N+1,\mu+N-1\rangle\right]\;,
\end{eqnarray}
where $C'=\sqrt{\alpha^2+\gamma^2}$ and $C''=\sqrt{\beta^2+\gamma^2}$, can be prepared by  detecting
the $l$-th atom ($l\neq k$) either in the state $|g_l\rangle$
or in the state $|f_l\rangle$.

\section{Conclusions}

\indent

In conclusion, we show how different classes of entangled states, namely, 
two-qubit, three-qubit as well as two-qutrit entangled states can be prepared 
either by a deterministic way or in a probabilistic way. 
We use a collection of atoms resonantly interacting with a two-mode cavity. We show that the
entangled states can be derived from the null eigenstate 
(which is a multi-atom and bimodal field entangled state) of the relevant Hamiltonian.  
We discuss how two 
genuine inequivalent classes of tripartite entangled states like GHZ states and W-states can be prepared.
Further, we have prepared the entangled states of two qutrits which are
two cavity modes each with three possible Fock states in this case.

\end{document}